 \documentclass[final,5p,times,twocolumn]{elsarticle}

\usepackage{amsmath,amssymb,amsfonts}
\usepackage{mathrsfs}
\usepackage{chemformula}
\usepackage{color}
\usepackage[normalem]{ulem}
\usepackage{stmaryrd}
\usepackage{diagbox}

 \biboptions{sort&compress}

\newcommand{\beq}{\begin{equation}}
\newcommand{\eeq}{\end{equation}}
\newcommand{\pfr}[2]{\ensuremath{\frac{\partial #1}{\partial #2}}}
\newcommand{\pfi}[2]{\ensuremath{{\partial #1}/{\partial #2}}}
\newcommand{\ep}{\varepsilon}
\newcommand{\Pec}{\textit{Pe}}
\newcommand{\Sch}{\textit{Sc}}

\newcommand{\eff}{\mathrm{eff}}

\newcommand{\mb}[1]{\mathbf{#1}}

\newcommand{\ob}[1]{\overline{#1}}

\newcommand{\mc}[1]{\mathcal{#1}}

\journal{Progress in Scale Modeling, an International Journal}

\begin{document}

\begin{frontmatter}

\title{Taylor dispersion in variable-density, variable-viscosity pulsatile flows}

\author{Prabakaran Rajamanickam$^{a*}$ and Adam D. Weiss$^b$}
\address{$^a$Department of Mathematics, University of Manchester, Manchester M13 9PL, United Kingdom\\
$^b$ATA Engineering, Inc., San Diego 92128, USA\\
Email:prabakaran.rajamanickam@manchester.ac.uk}

\begin{abstract}
The phenomenon of Taylor or shear-induced dispersion of a non-passive scalar field in a pulsatile pipe flow is investigated, accounting for the scalar field's influence on fluid density and transport coefficients. By employing multiple scale analysis, an effective one-dimensional, unsteady mixing problem for the scalar field is obtained, which includes the diffusion coefficient for shear-induced dispersion. The resulting governing equations are applicable to a range of scalar transport problems in pulsatile pipe flows.
\end{abstract}


\end{frontmatter}

\section{Introduction}
\label{sec:intro}

The dispersion of a passive scalar, such as a solute, in a steady, strong Poiseuille flow is governed by Taylor dispersion, also known as shear-induced dispersion, as first described by Taylor in the 1950s~\cite{taylor1953dispersion,taylor1954conditions}. The analogous effect in oscillatory flows was analysed by Watson~\cite{watson1983diffusion}, who provided explicit formulas for the effective diffusion coefficients in simple geometries. Independently, Prigozhin~\cite{prigozhin1982solute} reached similar findings. Zeldovich~\cite{zel1982exact} developed a similar result for  unidirectional periodic flows, a result that can be generalized to three-dimensional spatio-temporal periodic flows~\cite{majda1999simplified,rajamanickam2023effective}. The general conclusion drawn from these studies is that at low frequencies, the effective diffusion approximates shear-induced dispersion for steady flows, while at high frequencies, it becomes negligible. For instance, in an oscillatory pipe flow operating at very high frequencies, the flow becomes nearly uniform across the cross-section, confining shear to thin boundary layers at the pipe wall, thereby rendering shear-induced dispersion negligible.

While extensive research has explored variations of the original problem, the impact of hydrodynamic influences on Taylor dispersion remains relatively unexplored.  Previous research has investigated the impact of variable density in steady Poiseuille flows in~\cite{oltean2004transport,felder2004dispersion} and, more recently, in the context of combustion, in~\cite{pearce2014taylor,rajamanickam2021effects,rajamanickam2023thick,rajamanickam2024effect}. The combined effects of variable density and viscosity have been considered in a recent study~\cite{rajamanickam2025shear} on buoyancy-induced dispersion, a notion originally introduced by Erdogan and Chatwin~\cite{erdogan1967effects}. 

Hydrodynamic influences emerge when the scalar field, rather than being passive, actively alters fluid density and viscosity. The resulting fluid motion, in turn, modifies the scalar field's dispersion. In this paper, we investigate these influences within a pulsatile pipe flow. This analysis will culminate in an effective one-dimensional, unsteady mixing problem for the scalar field, incorporating the shear-induced diffusion coefficient.

\section{Problem formulation}
\label{sec:formulation}

Let us introduce a compact distribution of a non-passive scalar field, $c=c(\mb x^*,t^*)$, into a fluid in an infinitely long pipe of radius $a$, exhibiting pulsatile motion driven by the longitudinal pressure gradient
\begin{equation}
    \pfr{p^*}{z^*} = - G -  \widetilde{G} \cos \omega t^* ,  \label{prgr}
\end{equation}
where $G$ and $\widetilde{G}$ represent the magnitudes of the steady and oscillating components of the pressure gradient, respectively, and $\omega$ is the frequency of oscillations. Let $l_m$ denote the characteristic length scale of mixing and  $t_m = l_m^2/D_\infty^*$ the corresponding diffusion time scale, where $D_\infty^*$ is the reference diffusion coefficient of the scalar. Since $c$ is not passive, it will impact the flow by altering the fluid density and viscosity. The fluid motion associated with these changes will in turn, affect the dispersion of the scalar field. This coupling mandates a concurrent analysis of the scalar transport and hydrodynamic equations. In this paper, we quantify this coupling in the Taylor-dispersion limit, which is
\begin{equation}
    \ep = \frac{a}{l_m}=\sqrt{\frac{t_a}{t_m}}\ll 1, \quad \Pec=\frac{U a}{D_\infty^*} \sim 1 \quad \text{and}\quad \widetilde{\Pec}=\frac{\widetilde{U} a}{D_\infty^*} \sim 1, \label{lim1}
\end{equation}
where $t_a=a^2/D_\infty^*$ is the radial diffusion time, $U=Ga^2/8\rho_\infty^*\nu_\infty^*$  and $\widetilde{U}=\widetilde{G}a^2/8\rho_\infty^*\nu_\infty^*$ are the mean velocities of the steady and oscillatory flows; $\rho_\infty^*$ and $\nu_\infty^*$ represent the reference density and kinematic viscosity, respectively. In addition to $\ep$ and the two Peclet numbers, the analysis will employ two more dimensionless numbers, the Schmidt number $\Sch$ and the non-dimensional number, $\beta$, which are defined by
\begin{equation}
    \Sch = \frac{\nu_\infty^*}{D_\infty^*}\sim 1, \qquad \beta = \frac{\omega a^2}{D_\infty^*}\sim 1 . \label{lim2}
\end{equation}
Following Taylor's original analysis~\cite{taylor1953dispersion,linan2020taylor,rajamanickam2021effects}, we adopt a frame moving with the speed $U$ ensuring that the mass flux, averaged over the cross-section and time period, is zero as $x\to \infty$. The moving coordinate is denoted by 
\begin{equation}
    x^*=z^*-Ut^*.
\end{equation}
For pulsatile flows with zero mean ($G=0$), no coordinate transformation is required.

To facilitate the analysis, introduce the following non-dimensional variables,
\begin{align}
    &t = \frac{t^*}{t_m}, \quad x = \frac{x^*}{l_m}, \quad r=\frac{r^*}{a},  \quad {\mb v} = \frac{{\mb v}^*l_m}{D_\infty^*}, \label{nondim1}  \\ &p = \frac{p^*a^2}{\rho_\infty^*\nu_\infty^*D_\infty^*}, \quad  \rho = \frac{\rho^*}{\rho_\infty^*}, \quad 
    \mu = \frac{\rho^* \nu^*}{\rho_\infty^*\nu_\infty^*}, \quad \lambda=\frac{\rho^* D^*}{\rho_\infty^* D_\infty^*}.\label{nondim2}
\end{align}
In terms of these variables, the low Mach-number Navier--Stokes equations can be written as
\begin{align}
    \pfr{\rho}{t} +  \nabla\cdot(\rho {\mb v}) &=0, \label{3dcont}\\
    \frac{\rho}{\Sch}\left(\pfr{{\mb v}}{t}+ {\mb v}\cdot \nabla{\mb v}\right) &= -\frac{\nabla p}{\ep^2} + \nabla\cdot\left\{ \mu\left[\nabla {\mb v} + (\nabla {\mb v})^{T}-\tfrac{2}{3}(\nabla\cdot{\mb v})\mathbf{I}\right]\right\} , \label{3dmom}\\
   \rho\left(\pfr{c}{t}+ {\mb v}\cdot \nabla c\right) &= \nabla\cdot(\lambda \nabla c) + Q(c), \label{thetadim}\\   
    \rho =\rho(c), \quad \mu &= \mu(c), \quad \lambda=\lambda(c). \label{eqnst}
\end{align}
where $\nabla=(\partial_x,\partial_r/\ep,\partial_\theta/r\ep)$ reflects the different scalings used for the radial and longitudinal coordinates. The term $Q=Q(c)$ accounts for any volumetric sources or sinks, such as chemical reactions. Equation~\eqref{eqnst} represents the equation of state and the constitutive relations for the transport coefficients.

Applying no-slip boundary condition at the pipe wall with small leakages of the scalar field results in
\begin{align}
    \mb v=-\frac{\Pec}{\ep}\,\mb e_x, \quad  -\frac{\lambda}{\ep}\pfr{c}{r} =    \ep S(c)\quad \text{at} \quad r= 1,  \label{flux}
\end{align}
where $-(\Pec/\ep)\mb e_x$ is the velocity of the wall in the moving frame and $S=S(c)$ is a prescribed function. As usual, the required solution is expected to behave well at $r=0$. The imposed pressure gradient along the $x$-axis is given by
\begin{equation}
    \pfr{p}{x} = -\frac{8\Pec}{\ep}- \frac{8\widetilde{\Pec}}{\ep}\cos \frac{\beta t}{\ep^2}, \quad \text{as} \quad x\to\infty.\label{imposed1}
\end{equation}
In this study, we shall look for axisymmetric solutions by assuming that all physical variables are independent of $\theta$ and the azimuthal velocity component is zero, that is to say, the velocity components can be written as $\mb v = (u,v,0)$.

\section{Multiple scale analysis}

\subsection{Perturbation procedure}

This section develops the asymptotic solution to the problem defined in the preceding section, under the limits specified in~\eqref{lim1}--\eqref{lim2}. The problem involves two distinct time scales, $t_m$ and $t_a$ with the oscillation frequency $\omega^{-1}$ being of the order of $t_a$ since $\beta\sim 1$. Such problems are amenable to multiple scale analysis, which involves introducing two time coordinates, namely, the slow time, $t$, representing time measured in units of $t_m$ and the fast time, $\tau\equiv t/\ep^2$, representing time measured in units of $t_a$. The time dependence in the imposed pressure gradient~\eqref{imposed1} is associated with $\tau$, leading to
\begin{equation}
    \pfr{p}{x} =  -\frac{8\Pec}{\ep}- \frac{8\widetilde{\Pec}}{\ep} \cos \beta \tau.
\end{equation}
In the multiple scale analysis, the time derivatives are transformed according to
\begin{equation}
    \pfr{}{t} \mapsto \pfr{}{t} + \frac{1}{\ep^2}\pfr{}{\tau}.
\end{equation}
Moreover, the problem does not require the introduction of two length scales. This is because, at small length scales, $(\mb x^*\sim a)$, the local flow approximates an incompressible Poiseuille-type flow, which do not directly depend on the longitudinal coordinate. Thus, all physical variables, say $\varphi$, are assumed to be of the form $\varphi = \varphi(x,t,r,\tau)$, in which $(x,t)$ are referred to as the large-scale or slow coordinates and $(r,\tau)$ as  the small-scale or fast coordinates.

The required solution is then sought in the form of a regular perturbation series
\begin{align}
  u = \frac{u_0}{\ep} +    u_1  + \cdots, &\quad  v = 0 +   v_1  + \cdots, \\
   p = \frac{p_0}{\ep} + p_1+ \cdots , &\quad
   c = c_0 + \ep c_1+\cdots.  
\end{align}
The expansion for density is given by $\rho=\rho_0+\ep\rho_1+\ep^2\rho_2+\cdots$, where 
\begin{equation}
    \rho_0=\rho(c_0), \quad \rho_1 = c_1 \frac{d\rho_0}{dc_0}, \quad \rho_2=c_2\frac{d\rho_0}{dc_0} + \frac{c_1^2}{2}\frac{d^2 \rho_0}{dc_0^2}, \quad \dots
\end{equation}
Similar expansions can be written down for the functions $\mu(c)$, $\lambda(c)$, $Q(c)$ and $S(c)$. Substituting the perturbation expansions into equations~\eqref{3dcont}--\eqref{imposed1} and collecting terms of different orders of $\ep$, a set of problems at successive orders is obtained, which are solved below. To facilitate further calculations, it is convenient to define the average of a physical variable, say $\varphi$, over the small-scale coordinates as
\begin{equation}
   \langle \varphi\rangle = \frac{\beta}{2\pi}\int_{\tau}^{\tau+2\pi/\beta} \int_0^1 \varphi\, 2 rdrd\tau. \label{avg}
\end{equation}

\subsection{Leading-order scalar field}
The leading-order equation for the scalar field $c$ is 
\begin{equation}
    \rho_0\pfr{c_0}{\tau} = \frac{1}{r} \pfr{}{r}\left(\lambda_0 r\pfr{c_0}{r}\right),  
\end{equation}
which is subject to the condition $ \pfi{c_0}{r}=0$ at $r=0,1$ and the periodicity condition in $\tau$. This problem is satisfied  if $c_0$ is independent of the small-scale variables. Thus, we have $c_0=c_0(x,t)$ and as a result, $\rho_0=\rho_0( x,t)$, $\mu_0=\mu_0(x,t)$, etc.

\subsection{Leading-order flow field and first-order correction to the scalar field}
The leading-order hydrodynamic equations and the first-order scalar equation are coupled and are given by
\begin{align}
    \pfr{\rho_1}{\tau}+ \pfr{}{x}(\rho_0 u_0) + \frac{\rho_0}{r}\pfr{}{r}( r v_1) = 0,  \label{cont11}\\
    \frac{\rho_0}{\Sch}\pfr{u_0}{\tau} = - \pfr{p_0}{x} + \frac{\mu_0}{r}\pfr{}{r}\left(r\pfr{u_0}{r}\right), \label{xmom0}\\
    0 = -\pfr{p_0}{r}, \\
    \rho_0\pfr{c_1}{\tau} + \rho_0 u_0 \pfr{c_0}{x} = \frac{\lambda_0}{r} \pfr{}{r}\left( r\pfr{c_1}{r}\right),  \label{c11}
\end{align}
subject to the boundary conditions $u_0+\Pec=v_1=\pfi{c_1}{r}=0$ at $r=1$, $\pfi{u_0}{r}=v_1=\pfi{c_1}{r}=0$ at $r=0$, the periodicity condition in $\tau$ and the imposed pressure gradient $\pfi{p_0}{x}=-8\Pec-8\widetilde{\Pec}\cos\beta\tau$ as $x\to\infty$. By integrating over the small-scale coordinates $(r,\tau)$, the continuity and the scalar equations yield
\begin{equation}
  \pfr{}{x}(\rho_0 \langle u_0 \rangle)=-\left\langle \pfr{\rho_1}{\tau} \right\rangle =   \langle u_0\rangle \frac{\rm d\rho_0}{\rm dc_0} \pfr{c_0}{x} \label{firstconst}
\end{equation}
which implies that 
\begin{equation}
    \pfr{\langle u_0\rangle }{x}=0 \quad \Rightarrow \quad \langle u_0 \rangle = 0
\end{equation}
since no net averaged mass flux is imposed in the moving frame as $x\to \infty$. In fact, the solvability condition for $v_1$ in the continuity equation~\eqref{cont11} requires not only $\langle u_0\rangle$ to be $x$-independent, but also the instantaneous cross-sectional average of $u_0$, i.e.,
\begin{equation} 
    \pfr{}{x}\int_0^r u_0 r dr = 0, \quad \Rightarrow \quad \int_0^r u_0 r dr = \frac{8}{\beta^2}\widetilde{\Pec}\Sch^2\mc Re\{A_\infty e^{i\beta \tau}\}, \label{radialavg}
\end{equation}
where the right-hand side is the value of $\int_0^r u_0 r dr$ evaluated at $x\to\infty$, where $\rho_0=\mu_0=1$; the complex constant $A_\infty$ is given below in~\eqref{constA}. However, it is also clear that at finite values of $x$, $u_0$ is expected to vary with  $x$ due to its dependence on $\rho_0$ and $\mu_0$. Consequently, to satisfy the above constraint, the flow will adjust itself by modifying the imposed pressure gradient  $-8\Pec-8\widetilde{\Pec}\mc Re\{e^{i\beta\tau}\}$ within the pipe.

Integrating~\eqref{cont11}-\eqref{radialavg}, we readily obtain
\begin{align}
    p_0 &= - 8\Pec \int^x\mu_0\,dx -8\widetilde{\Pec} \mc Re\left\{e^{i\beta\tau} \int^x\frac{\rho_0^2A}{\mu_0}dx\right\}, \label{solp0}\\
     u_0 &= \Pec (1-2r^2) + \frac{8\widetilde{\Pec}\Sch\rho_0}{\beta \mu_0 }  \mc Re\left\{i A f e^{i\beta \tau}\right\}  , \label{solu0}\\
    c_1 &= C + \frac{\Pec\rho_0}{8\lambda_0}\pfr{c_0}{x}(2r^2-r^4)+  \frac{8\widetilde{\Pec}\Sch\rho_0}{\beta^2\mu_0}\pfr{c_0}{x}\mc Re\left\{ A g e^{i\beta \tau}\right\}, \label{solc1}   
\end{align}
where $C=C(x,t)$ is the integration constant, 
\begin{equation}
    A(x,t)=  A_\infty\left(\int_0^{\alpha} f \eta d\eta\right)^{-1}, \label{varA}
\end{equation}
and the functions $f(\eta)$ and $g(\eta)$ with $\eta = \alpha r$, $\alpha=\sqrt{\beta\rho_0/\mu_0\Sch} $ satisfy the ordinary differential equations
\begin{align}
     (\eta f')'/\eta   = i(f+1), &\qquad f'(0)=f(\alpha)=0, \label{feq}\\
    \sigma^{-1}(\eta g')'/\eta  = i(f+g), &\qquad g'(0)=g'(\alpha)=0. \label{geq}
\end{align}
Here,  prime denotes differentiation with respect to $\eta$, and $\sigma(x,t)= \mu_0\Sch/\lambda_0=\nu_*/D_*$ is the variable Schmidt number. The solution for the first equation~\eqref{feq} is given by
\begin{equation}
    f(\eta) = \frac{I_0(\sqrt{i}\eta)}{I_0(\sqrt{i}\alpha)}-1, \label{fsol} 
\end{equation}
which involves the modified Bessel functions, $I_n$. From this, we obtain the formulas for $A$ and the constant $A_\infty$ in~\eqref{radialavg},
\begin{equation}
    A(x,t) = -\frac{2A_\infty I_0(\sqrt{i}\alpha)}{\alpha^2 I_2(\sqrt{i}\alpha)}, \quad A_\infty =-\frac{\beta I_2(\sqrt{i\beta/\Sch})}{2\Sch I_0(\sqrt{i\beta/\Sch})}. \label{constA}
\end{equation}
In the case of constant-density, constant-viscosity approximation wherein $\rho_0=\mu_0=1$, we have $A=1$. It is noteworthy that the component of $u_0$ pertaining to $\Pec$ maintains the form of a standard Poiseuille flow, whereas the component related to $\widetilde{\Pec}$ exhibits a significantly different structure and also evolves with the large-scale coordinates $(x,t)$. The pressure field $p_0$, however, varies on the large-scale coordinates in both cases. 

The solution for the second equation~\eqref{geq} is given by
\begin{align}
    g(\eta) &= 1 +  \frac{\sigma}{1-\sigma}\left[\frac{I_0(\sqrt{i}\eta)}{I_0(\sqrt{i}\alpha)} - \frac{1}{\sqrt{\sigma}}\frac{I_1(\sqrt{i}\alpha)}{I_0(\sqrt{i}\alpha)}\frac{I_0(\sqrt{i\sigma}\eta)}{I_1(\sqrt{i\sigma}\alpha)}\right], \quad  \sigma\neq 1, \label{gsol1}\\
     g(\eta)&=1 +  \frac{\sqrt{i}}{2}\left[\frac{\eta I_1(\sqrt{i}\eta)}{I_0(\sqrt{i}\alpha)}- \frac{\alpha I_0(\sqrt{i}\eta)}{I_1(\sqrt{i}\alpha)}\right], \quad \sigma=1. \label{gsol2}
\end{align}
These solutions for $f(\eta)$ and $g(\eta)$ were previously derived by Watson~\cite{watson1983diffusion} and Prigozhin~\cite{prigozhin1982solute}, where the definition of $\eta$ was simpler and $\sigma$ was constant in their constant-density, constant-viscosity models. In our case, $\sigma=\sigma(x,t)$, and the apparent singularity at $\sigma=1$ in the first formula is only transitional; therefore, no issues arise if $\sigma$ approaches unity at some spatial point or at a particular time.

The solution for $v_1$ will not be required for the subsequent analysis and will not be discussed further. Moreover, the solvability of $v_1$ has already been ensured in~\eqref{radialavg}.

\subsection{Governing equations at the subsequent order}

The governing equations at the next order of the perturbation expansion are the first-order hydrodynamic equations and the second-order scalar equation, which are given by
\begin{align}
     &\pfr{\rho_0}{t} + \pfr{\rho_2}{\tau} + \pfr{}{x}(\rho_0 u_1+\rho_1 u_0 ) =-\frac{1}{r}\pfr{}{r}(\rho_0 r v_2 + \rho_1 r v_1) , \label{cont2}\\
     &\frac{1}{\Sch} \left(\rho_0\pfr{u_1}{\tau} +\rho_1\pfr{u_0}{\tau}\right) + \frac{\rho_0}{ \Sch} \left( u_0 \pfr{u_0}{x} + v_1 \pfr{u_0}{r}\right) \nonumber \\  &= -\pfr{p_1}{x} + \frac{1}{r} \pfr{}{r} \left(\mu_0 r\pfr{u_1}{r}\right),\\
      &0= - \pfr{p_1}{r}, \label{rmom2}\\
      &\rho_0 \pfr{c_0}{t} + \rho_0 \pfr{c_2}{\tau}  + (\rho_0 u_1 + \rho_1 u_0)\pfr{c_0}{x} + \pfr{}{\tau}(\rho_1 c_1) +  \pfr{}{x}(\rho_0 u_0 c_1) \nonumber \\ & + \frac{1}{r}\pfr{}{r}(\rho_0 rv_1c_1) 
      =\pfr{}{x}\left(\lambda_0\pfr{c_0}{x}\right)+ \frac{1}{r}\pfr{}{r}\left(\lambda_0 r\pfr{c_2}{r}\right) + Q_0, \label{conc2}
\end{align}
which are subject to the conditions $u_1=v_2=\lambda_0 \pfi{c_2}{r}+S_0=0$ at $r=1$, $\pfi{u_1}{r}=v_2=\pfi{c_2}{r}=0$ at $r=0$ and the periodic condition in $\tau$. Inspection of these equations reveals that the solution for $c_2$ will include a steady part (but still varying on the slow time $t$) and an oscillatory part with frequencies $\beta$ and $2\beta$. Based on this and the solution for $c_1$, we can infer that
\begin{equation}
    \left\langle \pfr{c_2}{\tau}\right\rangle = \left\langle \pfr{}{\tau}(\rho_1 c_1) \right\rangle= \left\langle \pfr{\rho_2}{\tau}\right\rangle =0.
\end{equation}

\subsection{Averaged, one-dimensional governing equations}

By averaging the governing equations over the small-scale coordinates, we aim to obtain a one-dimensional description of the mixing process. By inspecting the continuity equation~\eqref{cont2}, we can define the effective mass flux $\rho_0 u_\eff$ by~\cite{rajamanickam2021effects,rajamanickam2024effect}
\begin{equation}
    \rho_0 u_\eff = \langle \rho_0 u_1 + \rho_1 u_0\rangle. \label{masseff}
\end{equation}
Averaging of the continuity equation~\eqref{cont2} and the scalar equation~\eqref{conc2} yields a set of equations that can be cast into a one-dimensional unsteady mixing problem, similar to those described in~\cite{lin1976asymptotic,rajamanickam2021effects}. Averaging of the convective term $\pfi{(\rho_0u_0 c_1)}{x}$ will result in a flow-induced diffusion term.

Dropping the subscript ``$0$" for clarity in the averaged equations, we arrive at
\begin{align}
    \pfr{\rho}{t} +   \pfr{}{x}(\rho u_\eff )  &= 0, \label{finalcont}\\
    \rho \pfr{c}{t} +  \rho u_\eff \pfr{c}{x}    
    &= \pfr{}{x}\left(\rho D_\eff\pfr{c}{x}\right) +Q(c)-S(c), \label{finalc}\\
    \rho=\rho(c), \quad \mu &= \mu(c), \quad \lambda=\lambda(c),
\end{align}
where the effective diffusion coefficient $D_\eff$ is defined by
\begin{align}
     \rho D_\eff &= \lambda + \frac{\Pec^2\rho^2}{48\lambda}+ \frac{32\widetilde{\Pec}^2\rho^2}{\lambda}\frac{\Sch^4|A|^2}{\beta^4\sigma}\int_0^{\alpha}i  (\ob f g - f \ob g)\eta d\eta,\\
    &=\lambda  + \frac{\Pec^2\rho^2}{48\lambda} +\frac{64\widetilde{\Pec}^2\rho^2}{\lambda}\frac{\Sch^4|A|^2}{\beta^4\sigma^2} \int_0^{\alpha}  |g'|^2\eta d\eta.  \label{rhod}
\end{align}
The second representation follows from the first by eliminating $f$ using~\eqref{geq} and then by integrating by parts. This representation also clearly demonstrates the positive definiteness of $D_\eff$. The three terms for the diffusion coefficient, respectively, corresponds to molecular diffusion, Taylor diffusion for steady Poiseuille flow, and shear-induced diffusion for oscillatory pipe flow. 

It is worth noting that the Taylor dispersion term for the steady component depends on $\rho$ and $\lambda$, but not on $\mu$. This is not completely surprising, since 
\begin{equation}
    \mc Pe(x,t) = \frac{\rho\Pec }{\lambda}
\end{equation}
can be interpreted as a variable Peclet number $\mc Pe$ determined by the local values of the fluid density and the scalar diffusion coefficient, so that, when the imposed pressure gradient is steady $(\widetilde G=0)$, we have
\begin{equation}
     \rho D_\eff = \lambda \left(1 + \frac{\mc Pe^2}{48}\right).
\end{equation}
Therefore, when $d\rho/dc>0$ (e.g. saline water), accounting for the dependence $\rho=\rho(c)$ increases the effective Peclet number, whereas when $d\rho/dc<0$ (e.g. thermal mixing), it decreases the effective Peclet number. 

The third term in~\eqref{rhod}, pertaining to the oscillatory component, does not have a simple representation and depends on $\mu$ as well. A straightforward way to evaluate the integral in the formula for $D_\eff$ is to use the third and interesting representation devised by Watson~\cite{watson1983diffusion}. Writing $(\eta f')'/\eta \equiv \nabla^2f$ for convenience, we can deduce from~\eqref{feq}-\eqref{geq} that
\begin{equation}
    \ob g \nabla^2 f - f \nabla^2 \ob  g  - \nabla^2 \ob f+\sigma^{-1}\nabla^2\ob g  = (1+\sigma) if\ob g + i \sigma |f|^2 + i
\end{equation}
whose real part provides $i (1+\sigma) (\ob f g - f \ob g) = ( f\nabla^2\ob g - \ob g \nabla^2 f ) + (\ob f \nabla^2 g-g\nabla^2\ob f  ) + (\nabla^2 f + \nabla^2\ob f) - \sigma^{-1} (\nabla^2 g + \nabla^2 \ob g)$. Integrating this over the pipe cross-section with the boundary condition that $g'\equiv \pfi{g}{\mb n}=0$ at the pipe wall ($\eta=\alpha$) and using Green's identities, we obtain
\begin{align}
    (1+\sigma)\int_0 ^\alpha i(\ob f g - f \ob g) \eta d\eta  &= \frac{1}{2\pi }\int_0^{2\pi} \left[(1-\ob g)\pfr{f}{\mb n} + (1-g)\pfr{\ob f}{\mb n}\right]d\theta \\ &= 2\mc Re\left\{\left[(1-g)\ob f'\right]_{\eta=\alpha}\right\}.
\end{align}
Simplifying the above expression using~\eqref{fsol} and~\eqref{gsol1}, we finally obtain
\begin{align}
     \rho &D_\eff = \lambda + \frac{\Pec^2\rho^2}{48\lambda} \nonumber \\ &+ \frac{64\widetilde{\Pec}^2\rho^2\Sch^4|A|^2}{\lambda \beta^4(1-\sigma^2)}\mc Re\left\{\left[\frac{1}{\sqrt{\sigma}}\frac{I_1(\sqrt{i}\alpha)}{I_0(\sqrt{i}\alpha)}\frac{I_0(\sqrt{i\sigma}\alpha)}{I_1(\sqrt{i\sigma}\alpha)}-1\right]\frac{\sqrt{-i} I_1(\sqrt{-i}\alpha)}{I_0(\sqrt{-i} \alpha)}\right\},
\end{align}
where $\alpha=\sqrt{\beta\rho/\mu\Sch}$, $\sigma = \mu \Sch/\lambda$ and
\begin{equation}
     |A|^2 = \frac{\beta^2}{\alpha^4 \Sch^2} \frac{|I_0(\sqrt{i}\alpha)|^2}{|I_2(\sqrt{i}\alpha)|^2} \frac{|I_2(\sqrt{i\beta/\Sch})|^2 }{|I_0(\sqrt{i\beta/\Sch}) |^2}.
\end{equation}

To determine the effective mass flux $\rho u_\eff$~\eqref{masseff}, we need to solve~\eqref{cont2}--\eqref{rmom2}. However, in one-dimensional, unsteady problem, the expression for $u_\eff$ serves to determine the pressure field $p_1$ and is not directly required for analysing the scalar equation itself. The continuity equation~\eqref{finalcont} and $u_\eff$ are eliminated by introducing the Lagrangian coordinate~\cite{zel2002physics,von2004mathematical}, 
\begin{equation}
    \psi = \int_0^x \rho dx,
\end{equation}
such that $\rho = \pfi{\psi}{x}$ and $\rho u_\eff = - \pfi{\psi}{t}$. The equation~\eqref{finalc} then simplifies to
\begin{equation}
    \pfr{c}{t} = \pfr{}{\psi}\left(\rho^2D_\eff\pfr{c}{\psi}\right) + \frac{1}{\rho}(Q-S) \label{finaleq}
\end{equation}
which can be solved with appropriate initial and boundary conditions.

\bibliographystyle{elsarticle-num}
\bibliography{elsarticle-template}

\end{document}